\documentclass{article}
\usepackage[english]{babel}
\usepackage[utf8]{inputenc}
\usepackage{johd}

\title{A perspective on the use of health digital twins in computational pathology}

\author{Manuel Cossio$^{a}$$^{*}$\\
        \small $^{a}$Dept. of Mathematics and Computer Science, Universitat de Barcelona, Barcelona, Spain. \\
       
        \small $^{*}$Corresponding author: Manuel, Cossio; \tt{manuel.cossio@ub.edu}
}

\date{} 

\begin{document}

\maketitle

\begin{abstract} 
\noindent A digital health twin can be defined as a virtual model of a physical person, in this specific case, a patient. This virtual model is constituted by multidimensional data that can host from clinical, molecular and therapeutic parameters to sensor data and living conditions. Given that in computational pathology, it is very important to have the information from image donors to create computational models, the integration of digital twins in this field could be crucial. However, since these virtual entities collect sensitive data from physical people, privacy safeguards must also be considered and implemented. With these data safeguards in place, health digital twins could integrate digital clinical trials and be necessary participants in the generation of real-world evidence, which could positively change both fields.   \end{abstract}

\noindent\\

\section{Introduction}

Thanks to advances in storage capabilities and the digitization of routine practices, healthcare is changing its operating paradigm  (\citealp{palabindala2016adoption}; \citealp{xiao2018opportunities}). Day after day, considerable amounts of data relating to the medical care of patients in hospitals and clinics are being stored. In fact, there are systems as large as regions and countries that operate mainly using electronic health records (EHR) \citep{cossio2022rwd6}. These EHRs can store anything from text entries in defined variables to unstructured text with medical images. Therefore, it is essential to be able to have some entity that centralizes all the data per patient\citep{tayefi2021challenges}. This entity could bear some resemblance to the actual patient and be updated as new studies are performed. In fact, the humanization of the data also allows the addition of other types of information such as housing conditions and access to basic services. In this brief article, some particular aspects of these centralized and humanized entities will be discussed.

\section{What is a health digital twin?}

A health digital twin (HDT) could be defined as a virtual space model of a patient. This virtual model defines its shape and contour using multimodal clinical data, population data, socio-economic data, behavioral data and real-time data that may come from wearable sensors (Figure \ref{datatypes}). The HDT can be highly multidimensional if constructed using all available variables. On the other hand, its dimensions can be reduced by selecting the variables to construct it. The latter is often the case when HDTs are built for a particular purpose (\citealp{venkatesh2022health}).

\begin{figure}[H]
\centering
\includegraphics[width=1\textwidth]{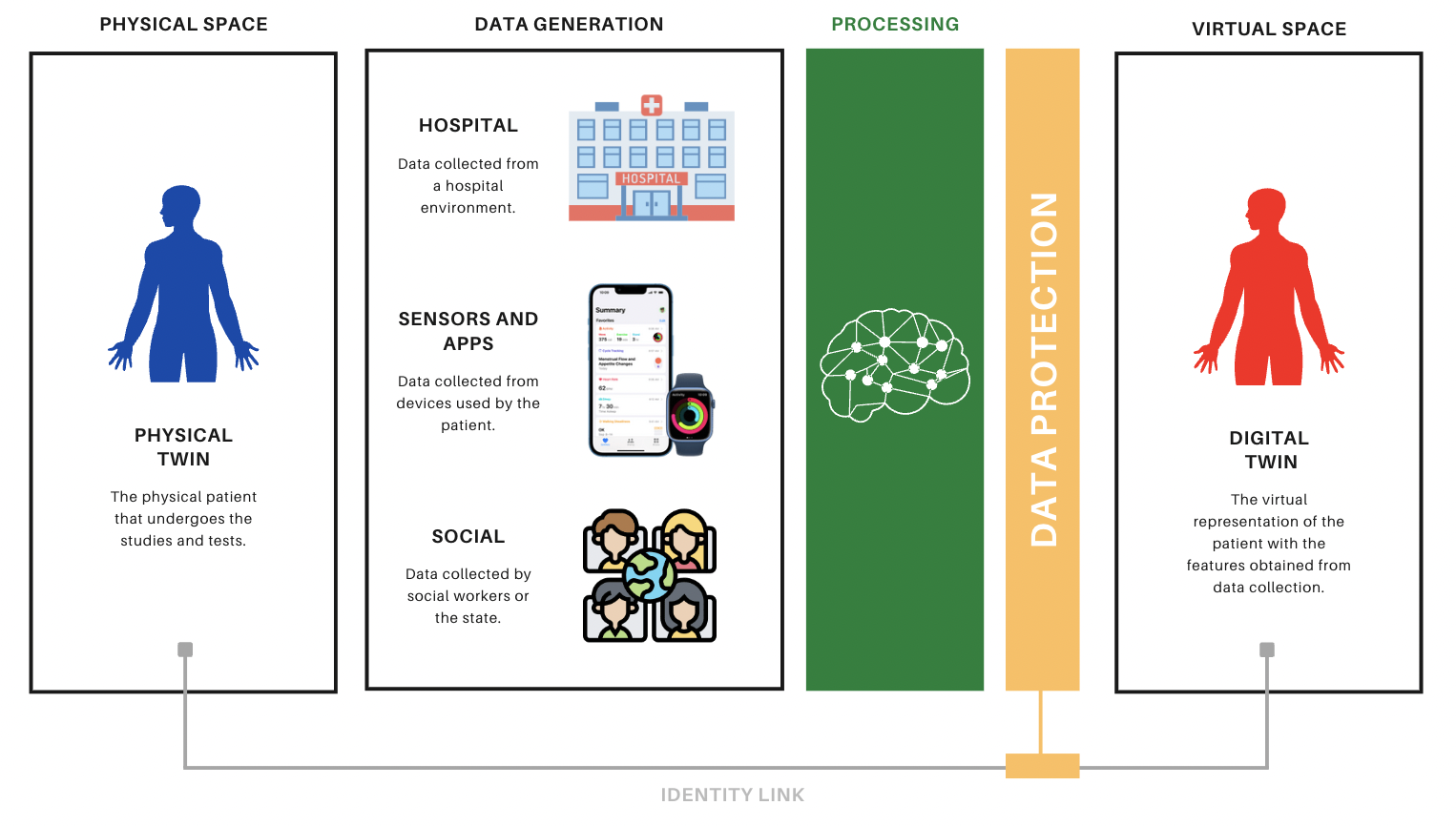}
\caption{\label{GenerationHDT} \textbf{Representation of the generation process of a health digital twin (HDT).} Processing includes data mining and application of artificial intelligence algorithms for automatization.  }
\end{figure}

\section{Evolution of digital twins for healthcare applications}

According to \citet{coorey2022health}, the term HDT comes from an adaptation of a similar term used in industrial engineering to refer to virtual models of physical systems that can be tested virtually. These virtual tests were used to refine the settings before manufacturing the devices and thus save time and resources in fine-tuning \citep{thuemmler2017health}. In fact, prototypes could also be manufactured in small numbers and sensors could be added. These sensors could record variables during the ride and improve the digital twin. The improved digital twin could be the target of virtual tests that were more reliable of the physical phenomena to which it was subjected (\citealp{coorey2022health}; \citealp{thuemmler2017health}).
The digital twin concept was, after what was explained in the previous paragraph, derived to healthcare. Within this field, we have numerous applications where data were integrated to build these virtual entities. In this way, we have some research works with HDT in cardiology (\citealp{coorey2022health}; \citealp{mazumder2019synthetic}; \citealp{gillette2021framework}; \citealp{herrgaardh2022digital}), public health \citep{kamel2021digital}, oncology (\citealp{zhang2020cyber}; \citealp{kaul2022role}; \citealp{meraghni2021towards}), immunology \citep{laubenbacher2022building}, neurology \citep{voigt2021digital}, infectology \citep{laubenbacher2021using} and nutrition \citep{gkouskou2020virtual}.

\section{HDTs environment}
Among the components that make up and should make up the environment of an HDT we mention some of them. 

\subsection{Artificial intelligence}
We can include within this category, all those computational components that collaborate in the emulation of human reasoning (Figure \ref{GenerationHDT}). These components are the ones that will provide the system with different degrees of autonomy, reactivity and self-initiation. \citet{coorey2022health} includes in this category entities such as:

\begin{itemize}
    \item Computational power
    \item Big data processing
    \item Machine learning
    \item Pattern recognition
\end{itemize}

We could go deeper in this listing and add possible interesting partners:

\begin{itemize}
    \item Computer vision
    \item Multiagents systems
   
\end{itemize}

\subsection{Internet of things (IoT)}
IoT refers to the physical and virtual means by which data can be exchanged \citep{coorey2022health}. Data may flow from inputs to servers, which may be housed in different physical locations (Figure \ref{GenerationHDT}). 

\subsection{Data exchange}

Between the HDT and the physical person (physical twin) there must be a constant exchange of information. This information can be acquired from the clinical and molecular study of the physical person and from the measurement of its behavior (sensoring)\citep{coorey2022health}. The closer the gap between these two entities, the more reliable the virtual model (HDT) will be.

\subsection{Data privacy and protection}
In this brief paper, we add this fourth component, which we believe is very important to include. There should be between the HDT and the physical person, a mechanism to protect privacy and safeguard anonymity. The HDT contains a lot of sensitive information. Only the parties necessary to carry out a diagnosis or emulation of a therapy should have access to it, with the prior consent of the individual. To prevent unauthorized third parties from gaining access to the HDT, the HDT information could be stored in a decentralized form with some form of encryption. Above all, the identity link between the physical person and the virtual model (Figure \ref{GenerationHDT}) would have to have several levels of protection. Some ideas in this regard will be developed in future sections (Section \ref{security}).

\section{Data: where it can come from?}\label{data-origin}

\begin{figure}[H]
\centering
\includegraphics[width=1\textwidth]{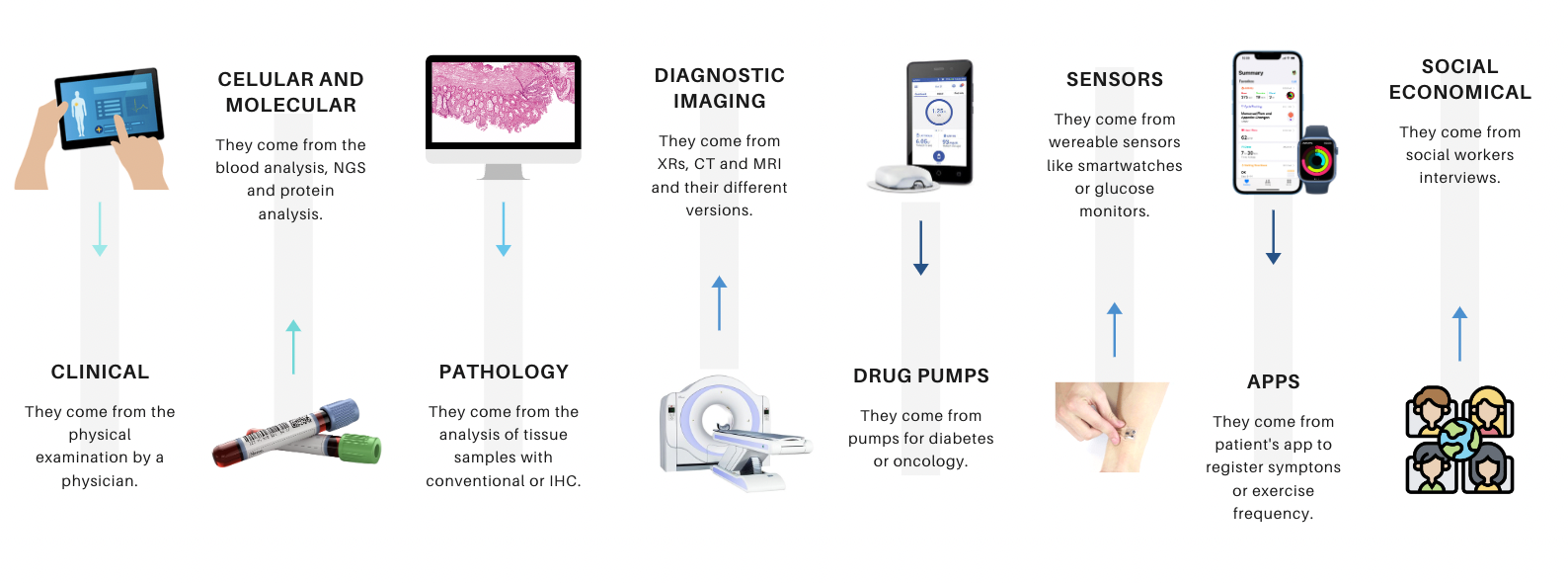}
\caption{\label{datatypes} \textbf{Different sources of data that can conform a health digital twin (HDT).} IHC, immunohistochemistry; XR, X rays; CT, computed tomography; MRI, magnetic resonance imaging; }
\end{figure}

Thanks to the incorporation of new diagnostic techniques and the increasing digitization of medical records, the data (Figure \ref{datatypes}) that can make up an HDT are extensive:

\begin{itemize}
    \item Clinical: They are obtained by the physician when performing the physical examination and asking about signs and symptoms. They can be in the form of free notes or in categories \citep{panahiazar2015using}. 
    \item Celular and molecular: they involve the prescription of laboratory tests. These tests can range from routine blood chemistry (such as hematocrit) or specific techniques such as gene, exome or genome sequencing. Sequencing can be on panels for diagnostic genes for specific pathologies or the patient's entire exome or genome \citep{panahiazar2015using}. 
    \item Pathology: they involve digitizing the tissue slides at multiple levels of magnification. The slides can correspond to classical pathology with conventional stains (e.g. hematoxylin and eosin) or to molecular pathology such as IHC which detects specific cellular markers \citep{feng2020deep}. 
    \item Diagnostic imaging: they involve imaging of internal organs and tissues. They may be with high-energy radiation, such as XR or CT, or by means of a high-intensity magnetic field, such as MRI. The first two should be used as little as possible and in justified cases, since radiation is harmful to patients in high doses\citep{pesapane2022digital}. 
    \item Drug pumps: involved in the dispensing of medicines to patients. These data involve amount of medication administered, frequency and treatment plan. Automatic pumps can dispense insulin, chemotherapy agents, antibodies for immunotherapy and enzymes for enzyme replacement therapy \citep{swapna2021diabetes}. 
    \item Sensor: they involve real-time measurement of the patient's vital signs such as heart rate, blood oxygenation, movement and degree of physical activity. They can also monitor sleep quality and some of them, of greater complexity and prior implantation, can monitor serum glucose levels \citep{mauldin2018smartfall}. 
    \item Apps: they involve the collection of patient data through interaction with an interface. These data can be level of drowsiness, type of food eaten, amount of food or state of pain. Some apps can also perform small tests, such as visual acuity tests, to collect data on sensory impairment \citep{brady2015smartphone}. 
    \item Social-economical: involve the collection of social and community data from the patient's environment. These data may include income level, access to essential services, disability, level of education achieved, interaction with the community where he/she lives, access to care services in sickness or old age. Some databases, such as the Secure Anonymised Information Linkage (SAIL) include such data, for example in its Index Multiple Deprivation (IMD) \citep{schnier2020secure}. 
\end{itemize}

\section{HDTs with computational pathology}

As we saw in Section \ref{data-origin}, part of the data that make up an HDT can be digital pathology images. Now, how can the use of HDT impact the development of computational pathology?

\subsection{Clinical and molecular profiling}

General clinical information on the patient's condition plus the blood and molecular marker profile is crucial to give context to what is found on the digital pathology images (Figure \ref{pathology}). For example, in a study by \citet{qu2021genetic} they were able to develop a deep learning model to predict point mutations (such as RB1 and CDH1) in whole slide images (WSIs) of breast cancer. The molecular profile data for each patient was obtained from EHR. The algorithm could be developed thanks to the availability of these data for each patient in the study.

\subsection{Diagnostic imaging}

Information provided by medical imaging studies, which may be contained in HDTs, can contribute positively to computational pathology. In a study by \citet{boehm2022multimodal}, the authors combined information from contrast-enhanced computed tomography (CE-CT) with that from WSI to address the diagnosis of high-grade serous ovarian cancer. The work included the construction of a multimodal model that segmented omental and adnexal disease on CE-CT and combined these features with tissue type and cell morphology from WSI analysis.

\subsection{Treatment plans}
Since HDTs can also store information on the type and degree of development of treatment plans, this information may also be important for the advancement of computational pathology. For example, in a study by \citet{lipkova2022deep} they were able to predict the type of cardiac allograft rejection in transplanted patients with various anti-rejection therapy regimens. The study used WSI from several cohorts of patients who had their therapeutic regimens recorded by automatic pumps in the EHRs. 

\subsection{Health apps}

In previous sections, we saw how mobile applications can help in the early screening of patients. Many of these apps can simply help with a signs and symptoms questionnaire. Some can perform cognitive or vision tests. Others, for example, can be key in accelerating early diagnosis of skin cancer by automatically analyzing mole photos by deep learning \citep{wibowo2020android}. With the score the patient obtains, he or she can then go to the hospital for a biopsy. The information from the app can be combined with that from the biopsy and then recorded in a WSI image and a trained algorithm can be used to detect tumor lesions\citep{hohn2021combining}.

\subsection{Tasks in computational pathology}

\begin{figure}[H]
\centering
\includegraphics[width=1\textwidth]{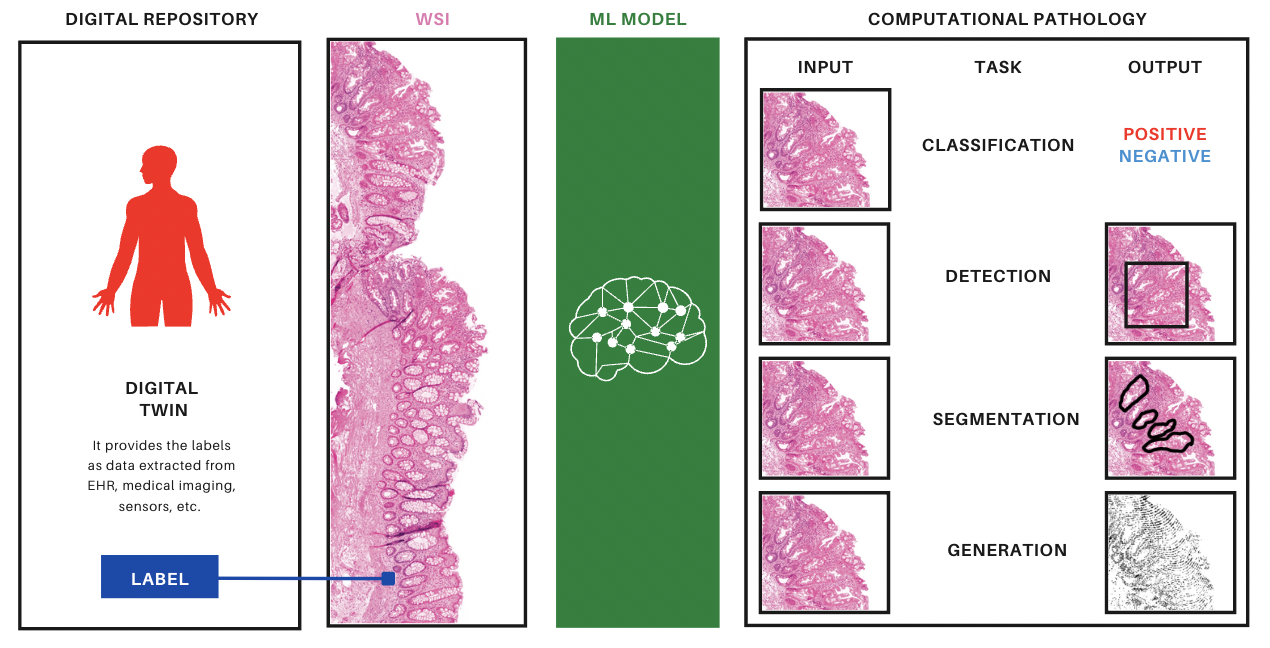}
\caption{\label{pathology} \textbf{Representation of health digital twins as label generators for computational pathology.} EHR, electronic health record; ML, machine learning; WSI, whole slide image.}
\end{figure}

There are many applications in computational pathology, especially given the large number of therapeutic areas that use pathology as part of the diagnostic and therapeutic process. Within computational pathology, four main tasks can be distinguished (Figure \ref{pathology}). The first is classification, where the input is an image and the output is a label (e.g., positive or negative). The second is detection, where the input is an image and the output is a bounding box with the detected region. This task allows not only to detect the region, but also to report its position. The third is segmentation, where the input is also an image and the output is a mask (a label for each pixel). The last one is the generation, where the input is an image and the output is another image similar to the input one but with a different style. This task is very useful in computational pathology to generate artificial stains on WSIs \citep{huo2021ai}.

\section{Digital clinical trials}

HDTs could have a considerable impact on clinical trials (CT). Given the increasing digitization of medical services and the performance of global CT for the same molecule, having HDTs for each patient would have a positive impact on data processing. Firstly, it would reduce operator bias since all the data could be processed in one place. Secondly, part of the processing could be done automatically using machine learning algorithms, which would speed up the obtaining of results. Third, the anonymity of the patients would be ensured since only the HDT would be available. Finally, if non-clinical (social) data are incorporated, the analysis could be broadened to determine if there is a component of the patients' environment that could be influencing the therapeutic response (\citealp{steinhubl2019digital}; \citealp{vamathevan2019applications}).

\section{Real world evidence (RWE)}

 Generation of RWE through the analysis of real world data (RWD) is crucial to enhance the therapeutic effect once the molecules have passed the CT stage. Since the CTs are subjected to the contingency that the patients are very typical and do not fully represent the real world population, having data from real patients would be optimal. Since the HDTs contain multidimensional information (laboratory, medical images, sensors, see Figure \ref{datatypes}) of very diverse patients, the analysis of these models could be key to obtain insights from real patients. In addition, since the HDTs may contain social and economic information, not only exclusively medical aspects but also multidisciplinary aspects that could be influencing the patient's recovery could be analyzed\citep{khosla2018real}. For example, if from the analysis of the HDTs parameters we observe that the variable access to emotional support and home care influences recovery after cancer treatment, this could be a factor to incorporate into the medical treatment plan (\citealp{lazar2018barriers}; \citealp{cossio2022rwd6}).

\section{Security and data protection}\label{security}

As mentioned before, HDTs contain a lot of sensitive information and are associated with the identity of individuals. The more extensive they are, i.e., the more features they contain, the higher the risk associated with their violation. Therefore, analyzing the possible ways to protect the information in the first place and the link with the individual in the second place is crucial to move towards secure practices. 

\subsection{Identifying risks}

As a first step, it is possible to identify parts of the programming of the models and pipelines that may have risk areas. A group in China worked on the development of a model that could identify parts of vulnerable code in HDTs built for lung cancer diagnosis. The model was trained with a ground truth that contained extensive pairs of vulnerable and non-vulnerable code from related applications \citep{zhang2020cyber}.  

\subsection{Decentralization and anonymization}

Helping to identify parts of the code that could be targeted is a good strategy. However, this is not enough to ensure that patients' identities are not compromised. The blockchain can be used to create a secure network to host and analyze DHTs. One could start with protecting the assembly of the DHT models by means of smart-contracts. These contracts could also regulate the users who would be authorized to upload data to the DHTs and those who would be authorized to read their information. To provide a greater degree of anonymity, large files (such as WSI, MRI or CT images) can be hosted on the DHTs through the use of interplanetary file system (IPFS) mapped with a content identifier (CID). In this way, part of the smart-contract would contain the identifiers separated from the IDs of physical persons and this would ensure that the data would be protected by one more barrier (\citealp{kumar2021integration}; \citealp{cossio2022ethereum}; \citealp{raj2021empowering}).

\section{Future perspectives}

In the previous paragraphs some important aspects concerning HDTs have been discussed very briefly. It has been shown the immensity of devices that can generate relevant data and the number of dimensions that can be accommodated by such an entity. It was also briefly mentioned the possible applications where the impact could be the greatest, such as clinical trials and RWE generation. However, we also looked at one aspect that needs to be addressed as quickly as possible, which is the protection of patient data. The data belong to the patients and therefore any application that is to be made using their data must have informed consent. In addition to consent, protection measures must be maximized and some strategies such as block chain or IPFS have been mentioned to generate a secure network that does not violate anonymization. It is this last part that has a promising future, because if holistic protection measures can be implemented, HDTs can be shared and studied worldwide. This is key to broaden the study populations of many pathologies, especially in the field of rare diseases. This field, as is well known, is characterized by too few patients to be able to design effective therapeutic strategies. In summary, HDTs will surely be able to provide many options for inclusive and representative studies that positively impact all human beings.

\bibliographystyle{johd}
\bibliography{bib}

\end{document}